\documentclass[%
 aip,
 amsmath,amssymb,
reprint,%
]{revtex4-2}

\usepackage{graphicx}
\usepackage{dcolumn}
\usepackage{bm}
\usepackage{mathptmx}
\usepackage{textcomp}
\usepackage{color}
\usepackage{orcidlink}

\newcommand{\beginsupplement}{%
        \setcounter{table}{0}
        \renewcommand{\thetable}{S\arabic{table}}%
        \setcounter{figure}{0}
        \renewcommand{\thefigure}{S\arabic{figure}}%
     }

\begin{document}


\title{Probing Ferroelectric Phase Transitions in Barium Titanate Single Crystals via $\it{in-situ}$ Second Harmonic Generation Microscopy}

\author{Benjamin Kirbus\, \orcidlink{0000-0002-8824-2244}}%
\affiliation{Institute of Applied Physics, Technische Universit\"at Dresden, N\"othnitzer Strasse 61, 01187 Dresden, Germany}
\affiliation{Leipzig Institute for Meteorology, Leipzig University, Stephanstrasse 3, 04103 Leipzig, Germany}
\author{Samuel D. Seddon\, \orcidlink{0000-0001-8900-9308}}
\affiliation{Institute of Applied Physics, Technische Universit\"at Dresden, N\"othnitzer Strasse 61, 01187 Dresden, Germany}
\author{Iuliia Kiseleva\, \orcidlink{0009-0002-5435-056X}}
\affiliation{Institute of Applied Physics, Technische Universit\"at Dresden, N\"othnitzer Strasse 61, 01187 Dresden, Germany}
\author{Elke Beyreuther\, \orcidlink{0000-0003-1899-603X}}
\email{elke.beyreuther@tu-dresden.de}%
\affiliation{Institute of Applied Physics, Technische Universit\"at Dresden, N\"othnitzer Strasse 61, 01187 Dresden, Germany}
\author{Michael R\"using\, \orcidlink{0000-0003-4682-4577}}%
\affiliation{Paderborn University, Integrated Quantum Optics, Institute for Photonic Quantum Systems (PhoQS), Warburger Str. 100, 33098 Paderborn, Germany}
\author{Lukas M. Eng\, \orcidlink{0000-0002-2484-4158}}
\email{lukas.eng@tu-dresden.de}%
\affiliation{Institute of Applied Physics, Technische Universit\"at Dresden, N\"othnitzer Strasse 61, 01187 Dresden, Germany}%
\affiliation{ct.qmat: Dresden-W\"urzburg Cluster of Excellence--EXC 2147, Technische Universit\"at Dresden, 01062 Dresden, Germany}%

\date{\today}

\begin{abstract}
Ferroelectric materials play a crucial role in a broad range of technologies due to their unique properties that are deeply connected to the pattern and behavior of their ferroelectric (FE) domains. Chief among them, barium titanate (BaTiO$_3$; BTO) sees widespread applications such as in electronics but equally is a ferroelectric model system for fundamental research, e.g., to study the interplay of such FE domains, the domain walls (DWs), and their macroscopic properties, owed to BTO's multiple and experimentally accessible phase transitions. Here, we employ Second Harmonic Generation Microscopy (SHGM) to \emph{in-situ} investigate the cubic-to-tetragonal (at $\sim$126$^\circ$C) and the tetragonal-to-orthorhombic (at $\sim$5$^\circ$C) phase transition in single-crystalline BTO via 3-dimensional (3D) DW mapping. We demonstrate that SHGM imaging provides the direct visualization of FE domain switching as well as the domain dynamics in 3D, shedding light on the interplay of the domain structure and the phase transition. These results allow us to extract the different transition temperatures locally, to unveil the hysteresis behavior, and to determine the type of phase transition at play (1st/2nd order) from the recorded SHGM data. The capabilities of SHGM in uncovering these crucial phenomena can easily be applied to other 
ferroelectrics to provide new possibilities for \emph{in-situ} engineering of advanced ferroic devices.
\end{abstract}

\maketitle

\section{\label{sec:Intro}Introduction}
Ferroelectric materials see widespread use for numerous technologies spanning a large range of applications, such as actuators and sensors, memories, capacitors, electro-optic modulators, or nonlinear-optical building blocks for usage in classical and nonlinear optics \cite{Wada2006,wessels2007ferroelectric,yoshino2009piezoelectric,Parizi2014,Zhao2020,Kampfe2020,Zhu2021,Han2022}. The characteristic feature of any ferroelectric is the spontaneous emergence of an electrical polarization $P_S$ in the absence of an external electric field. Areas of the same spontaneous polarization $P_S$ form ferroelectric domains, whose boundaries are referred to as domain walls (DWs). The domain structure and its patterning determine the macroscopic properties, e.g., its dielectric, piezoelectric, electro-optic, or nonlinear optical responses. Any application and device design therefore requires (i) a profound understanding of the domain structure and its DWs; (ii) knowledge on the impact of the domain structure on the (macroscopic) properties; and (iii) methods to systematically control and tailor the domain structure via so-called domain engineering \cite{Shur2015,Eng1999}. \\

The emergence of ferroelectricity is the product of a phase transition from the paraelectric phase of higher crystal symmetry to a ferroelectric phase of lower symmetry, when cooling the crystal below the transition temperature known as the Curie temperature $T_c$. In the formulation of Ginzburg-Landau-Devonshire theory -- which is broadly applied to describe ferroic ordering -- the order parameter associated with ferroelectricity is the spontaneous polarization $P_S$. The ferroelectric phase transition temperature and associated physical behaviors are highly dependent on subtle parameters such as sample history, doping, defects, or ambient conditions, and therefore is a crucial parameter to be understood when designing devices for applications. Furthermore, many material parameters like optical (linear, nonlinear), (di-)electrical, or mechanical (piezo-electrical) ones can experience gradual changes (e.g., second-order phase transitions), sudden jumps (e.g., first-order phase transitions) or any mixed behaviors when the temperature $T$ approaches $T_c$, as well as a large hysteresis during relaxation. Consequently, knowledge about the thermodynamic order of phase transitions and the characterization of parameters associated with the ferroelectric phase are of high relevance. The archetypal ferroelectric barium titanate (BaTiO$_3$, BTO), for example, exhibits multiple ferroelectric phase transitions of different order, which manifest in drastic changes to its microscopic (e.g., the domain structure) and macroscopic properties \cite{mer54,Potnis2011,Limboeck2014,Fu2015,doe18}. 

 Over the years, researchers have employed various techniques to explore the behavior of ferroelectric phase transitions. Often, such studies focus on macroscopic signals integrating over a large crystal volume, e.g. by employing X-ray diffraction, Raman spectroscopy, or investigations of other macroscopic properties connected to the crystallographic ordering \cite{ElMarssi2003}. To gain insight into the associated domain structures, such macroscopic studies do not provide the necessary spatial resolution. In this regard, some form of direct visualization technique with appropriate imaging speeds, spatial resolution, and sensitivity to domains or domain walls is required. Here, piezoresponse-force microscopy (PFM) is often considered the gold standard for analyzing ferroelectric domain structures, because it allows direct visualization of ferroelectric domains and domain walls with high spatial resolution (down to the nanometer-scale) \cite{gru97,Eng1998,Limboeck2014,doe18,kis18,sed24}. However, PFM is mostly sensitive to sample surfaces only, and hence limited to at most a micrometer of probing depth \cite{Johann2009,Roeper2024}. Furthermore, acquisition of a single image typically takes several minutes for images not exceeding 100~$\times$~100~\textmu m$^2$ in size, and thus can be considered relatively slow and locally restricted. Another common method to study domain structures upon phase transitions is polarized light microscopy (PLM), which allows a direct visualization of domains via the birefringence contrast. On the one hand there is a refractive index contrast between in-plane and out-of-plane domains, on the other hand a refractive index contrast also accompanies domain walls due to strain or incomplete screening \cite{hip50,Mulvmill1996,Wada1998,Golde2021}. Strictly speaking, this method provides a projected image of vertically stacked structures along the optical axis. Consequently, a measurable change in polarization requires a thick-enough crystal of several tens to hundreds of micrometers to generate a sufficient, detectable phase shift. Therefore, true 3D resolution is often not achievable. 

Contrasting the limitations outlined above, Second Harmonic Generation Microscopy (SHGM) enables fast (frame rates $> 1$~Hz), non-destructive, and high-resolution imaging of ferroelectric domain walls within 3D structures \cite{ruesing2019second,kir19,spychala2020spatially,spychala2020nonlinear,hegarty2022tun,Amber2022,Spychala2023}. SHGM makes use of the nonlinear optical effect of second-harmonic generation (SHG), where two photons at a pump wavelength are annihilated and a single photon at twice the frequency (half the wavelength) is generated. This process is very sensitive to underlying changes in crystal symmetry, allowing the visualization of both phase-transitions as well as local changes in the nonlinear susceptibility, which are seen at DWs \cite{cherifi2017non,cherifi2021shedding}. The SHG effect has been previously implemented to study phase transitions of BTO by integrating a \emph{macroscopic} signal over large crystal volumes \cite{Pugachev2012,Wang2018}, which, however, did not provide images of the domain structure. In this regard, SHGM offers a superior spatial resolution allowing to visualize the domain patterns. Due to its nature as a diffraction-limited scanning microscopy technique, voxel sizes down to 1~\textmu m$^3$ are possible in SHGM. By exploiting the inherent second-order nonlinear optical properties of ferroelectric materials, SHGM enables the direct observation of domain structures and their evolution, thereby offering insights into the dynamic processes during phase transitions \cite{Volker2006,Ayoub11,Ayoub17}. Recent developments in laser technology and microscope design promise SHGM with $> 1$~kHz frame rates with even higher temporal resolution \cite{Vittadello2021}.

In this work, we apply SHGM to investigate two phase-transitions and the associated domain structure \emph{in-situ} in the archetypal perovskite ferroelectric BaTiO$_3$. In particular, we study the paraelectric-ferroelectric I (cubic-tetragonal) transition and ferroelectric I-ferroelectric II (tetragonal-orthorhombic) phase transitions. The cubic-tetragonal phase transition temperature (Curie-temperature) will be labelled $T_{c1} \approx 126$~$^\circ$C here. Similarly, the tetragonal-orthorhombic transition will be referred to as $T_{c2} \approx 5$~$^\circ$C. Literature values for both transitions temperatures are scattered over a broad range of temperature values depending on sample history (hysteresis behavior), crystal stoichiometry, or microstructure \cite{Limboeck2014,Villafuerte2016,doe18,MAZUR2023}. 
We explore how SHGM imaging (i) can provide insights into the appearance of the 3D domain structure tens to hundreds of micrometers deep below the crystal surface with \textmu m resolution and (ii) allows to extract the phase transition temperatures $T_{c1}$ and $T_{c2}$ including the observation of hysteretic behavior. Furthermore, we discuss whether or not the thermodynamic order of the phase transition can directly be extracted from the SHGM data. 

\section{\label{sec:mat_meth}Methods and Materials}

\begin{figure}[ht]
\includegraphics[width=\linewidth]{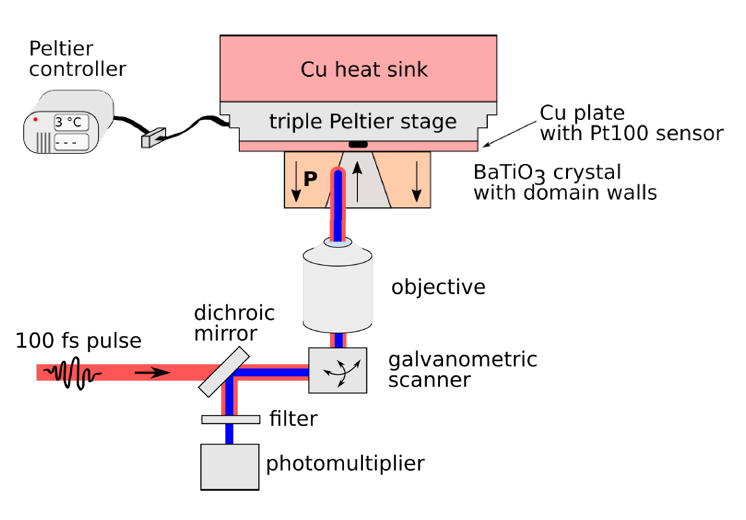}
\caption{\label{fig:setup} Sketch of the SHGM setup coupled with a triple Peltier heating/cooling stage. Here, $100$-fs laser pulses centered at approx. 900~nm wavelength are focused onto the barium titanate single crystals using an objective lens. Scanning is achieved via an xy-galvanometric scanning unit and the motorized 3D-stage, to which the sample and the Peltier stage are mounted. Due to the \v{C}erenkov-type phase matching, ferroelectric domain walls (DWs) efficiently convert the fundamental light beam to SHG light at 450~nm allowing 3D visualization of the DWs.}
\end{figure}

A schematic of the optical setup is drawn in Figure~\ref{fig:setup}. More details can be found in previous publications\cite{kaem14,kaem15,kir19,amber2021quantifying}. In this experiment, a titanium:sapphire laser (Mai Tai BB, Spectra Physics; pulse length $100$~fs; repetition rate $80$~MHz) is used in conjunction with a commercial laser scanning microscope (SP5~MP, Leica). The incident wavelength was $900$~nm with pulse energies of  $\sim 13$~nJ/pulse for the current study. The band gap of BaTiO$_3$ is  $\sim$ 3.1~eV (400~nm) \cite{Cox1966}; therefore a BTO sample is transparent for both pump wavelength (900~nm) and the SHG wavelength (450~nm) allowing for 3D imaging. The pump beam is focused onto the BTO sample using an air objective (NA~$0.70$, $20 \times$). The SHG signal is collected in reflection mode using the same objective lens and detected via photomultiplier tubes. The fundamental light is blocked by an appropriate edge filter (F75-680, Semrock). In this confocal laser scanning microscope, scanning in the x-y plane is carried out by fast galvanometric scanners. For depth scans along the sample's z-direction, the BTO crystal (and Peltier stage) is mounted on a 3D-motorized stage. Hence, typical image acquisition times for frames of around 400~$\times$~400~\textmu m$^2$ were approximately one second for the given parameters Typical full 3D scans with up to 100 slices recorded at different depth can be performed in $\sim$ 100 seconds.

The sample was mounted on top of a triple Peltier stage connected to a E5CN temperature controller (Omron Corp., Kyoto, Japan), as shown in Fig. \ref{fig:setup}. Temperatures from -10~$^\circ$C to +140~$^\circ$C were set and controlled with the help of a built-in Pt100 resistance thermometer. An accuracy of about $\Delta \mathrm{T} = \pm1$~K is specified by the manufacturer. 

For this study an undoped BaTiO$_3$ (100) epi-ready wafer (MaTeck GmbH, Jülich, Germany) with sample dimensions of $5 \times 5 \times 0.5~\mathrm{mm}^3$ was investigated. The sample's surface was pre-polished to optical quality to assure transparency for imaging. Naturally occurring domains were exploited, thus no additional poling procedure is required. The sample was not subject to any pre-annealing. 

The sample is repeatedly studied for multiple experimental runs above or below $T_{c1}$ or $T_{c2}$, respectively. To study the evolution of the domain structure during the phase transitions, images at a constant depth are taken with frame rates of $>1$~Hz. Based on the density and volume a heat capacity of $0.375$~J/K can be estimated for the sample \cite{Horowitz96}. Due to the slow cooling and heating rates ($<1$~K/s), the cooling power of the Peltier elements and the small heat capacity of the sample, each image is taken in (nearly) thermal equilibrium. The temperature at the Pt100 resistance thermometer is therefore considered to be the temperature of the sample within the error limit of the controller of $\pm 1$K. The incident laser and detector settings are kept constant during each experiment allowing to readily extract and compare intensity changes of the domain walls during each imaging run. 

\section{\label{sec:results_and_discussion}Results and discussion}

\subsection{SHGM at room temperature in BaTiO$_3$}

\begin{figure*}[ht]
\includegraphics[width=0.9\linewidth]{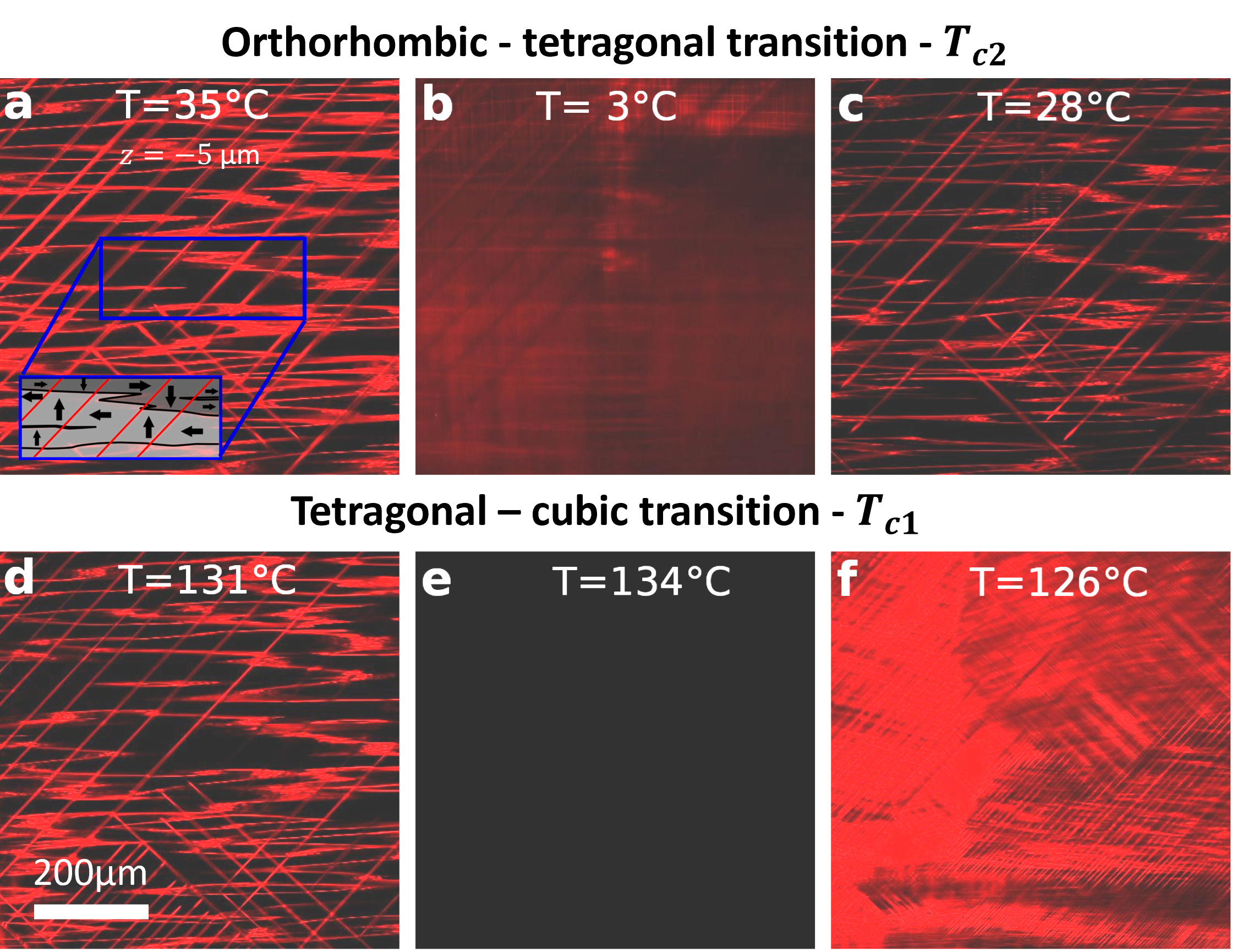}
\caption{\label{fig:domains} Near-surface domain pattern observed in the BTO sample at a focus depth of approximately 5~\textmu m. Here, the observed domain patterns in the SHG image correlate well with known domain patterns in BaTiO$_3$(100) with an selection of in-plane a-domains separated step-wise by 90$^\circ$ and 180$^\circ$ walls \cite{hip50,Limboeck2014}. Panels \textbf{a} to \textbf{c} show a typical behavior when cooling below $T_{c2}$ and relaxing back to room temperature. Overall, the same domain structure is seen indicating a strong influence of pinning for this transition. In contrast, when heating above the paraelectric-ferroelectric transition temperature $T_{c1}$ (panels \textbf{d} to \textbf{f}) a domain structure with a much higher density of domains emerges showing the result of an annealing cycle.}
\end{figure*}

SHG microscopy allows direct visualization of DWs in ferroelectrics. One example taken at room temperature close to the surface (at a depth of 5~\textmu m) is shown in Figs.~\ref{fig:domains}\textbf{a} to \textbf{f}, where domain walls are visible as bright lines. The contrast mechanism arising from DWs relative to the surrounding bulk is explained by two phenomena -- firstly, SHG within highly focused beams and far away from interfaces (be that the crystal surface or a DW) is fundamentally forbidden, which explains the dark background\cite{Spychala2023,hegarty2022tun}. Secondly, SHG at DWs is massively enhanced due to so-called \v{C}erenkov-type phase matching\cite{kaem14,kaem15,hegarty2022tun}. Our observation of this type of contrast is similar to previous studies on SHGM on BTO or lithium niobate crystals \cite{kaem15,bec22,kir19}. 

The observed structure in Fig.~\ref{fig:domains} correlates well with typical domain patterns in BaTiO$_3$(100), e.g., known from PLM or PFM  room temperature measurements\cite{Limboeck2014,gru97,hip50,Wada1998}. Here, close to the surface, DWs are seen which resemble the typical patterns of in-plane domains with 90$^{\circ}$ and 180$^{\circ}$ domain walls well known for (100)-cut BaTiO$_3$ \cite{Limboeck2014,gru97,hip50}. SHGM is usually not sensitive for the absolute domain orientation, i.e. distinguishing the absolute orientation of the spontaneaous polarization $P_S$ is not possible with standard SHGM and requires dedicated interference SHGM \cite{Yokota2021}. Therefore, the inset in Fig.~\ref{fig:domains}\textbf{a} represents only one plausible orientation of the polarization vectors, while the opposing orientations are also possible.

Apart from the \v{C}erenkov-type contrast at DWs other contrast mechanisms are known in ferroelectrics as well. First, the so-called interference contrast leads to dark-contrast at DWs \cite{ruesing2019second,spychala2020spatially,hegarty2022tun}, but is only appearing in very thin samples or close to a surface, and can, therefore, be neglected here, since imaging was performed sufficiently deep into the material. Second, effects of locally changed nonlinear susceptibility due to special DW sub-symmetries are typically weak, and require a dedicated polarization sensitive detection \cite{cherifi2017non,cherifi2021shedding,spychala2020nonlinear}. Consequently these are of minor relevance for this study. 

\subsection{The tetragonal-orthorhombic phase transition \label{sec:Tc2_results}}
 
As described above, when cooling BTO below a temperature of $T_{c2} \approx 5$~$^\circ$C, it undergoes a structural phase transition from the tetragonal to the orthorhombic phase. Figure~\ref{fig:domains}\textbf{a}-\textbf{c} shows three SHG images taken in a depth of 5~\textmu m: \textbf{a}, slightly above room temperature and before the phase transition; \textbf{b}, just below the phase transition temperature; and \textbf{c}, after relaxation back to room temperature. It is readily noted that when reaching the orthorhombic phase at $T < 3^{\circ}C$ the SHG intensity is vastly different compared to the case in panel~\textbf{a} and almost no signatures of DWs are visible. An unexpected result of cooling through the tetragonal-orthorhombic phase transition is that the SHG intensity within the orthorhombic phase (Fig.~\ref{fig:domains}\textbf{b}) only displays a weak, homogeneous SHG signal without any indications of DWs under the given laser power and detector settings. This is contrary to expectation as the orthorhombic crystal phase is also ferroelectric and thus should show a SHG response, as well as DWs. Further, the density of DWs (and number of domains) are likely to increase due to the increase from six to twelve possible domain orientations and due to the systematic splitting and twinning of domains\cite{Limboeck2014}. The incident laser polarization used in these measurements should be able to address at least some tensor elements of the nonlinear susceptibility tensor within orthorhombic domains. This weak SHG intensity is indicative of a much smaller nonlinear susceptibility tensor magnitude. This appears to be consistent with observations from the linear dielectric constants at low frequency, which are also smaller in the orthorhombic phase compared to the tetragonal phase \cite{Potnis2011,Fu2015}. After relaxation back to room temperature an almost identical domain structure with DWs at similar locations does appear. This hints at the strong role of pinning at defects for the position and place of domains during this transition. These measurements demonstrate clearly that there is a strong signature of the phase transition within SHGM intensity.

\subsection{The cubic-tetragonal phase transition \label{sec:Tc1_results}}

To probe the upper phase transition ($T_{c1} \approx 126 ^{\circ}C$), a similar experimental procedure was employed as in Section~\ref{sec:Tc2_results}, heating from $\sim 20^{\circ}C$ to $135^{\circ}C$ and then relaxing back to $30^{\circ}C$. During heating the domain pattern (Fig.~\ref{fig:domains}\textbf{d}) observed is identical to the pattern during the cooling measurements. The domain pattern overall stays the same up to the transition temperature (Fig.~\ref{fig:domains}\textbf{d}). As discussed below, only the total SHG intensity slowly decreases. When the temperature reaches the transition temperature at above $T > 131^{\circ}C$ all SHG intensity is lost and no intensity or DWs are detected (Fig.~\ref{fig:domains}\textbf{e}), indicating the structural phase transition to the cubic and paraelectric phase. When relaxing back to room temperature, a new domain pattern nucleates at  $\sim 126^{\circ}C$. Here, the overall domain pattern across the complete crystal shows the same shapes of in-plane domains with 90$^{\circ}$ and 180$^{\circ}$ domain walls as before, but a much higher density of domain walls, which form lamella-like patterns of approximately 5~\textmu m width separated by 180$^\circ$ walls. This indicates that whilst the original domain structure is stable before reaching the cubic-tetragonal phase transition, when cooling back to room temperature defects and pinning sites play a far less crucial role.

\subsection{Thermodynamic order of the phase transitions}

\begin{figure}
\includegraphics[width=\linewidth]{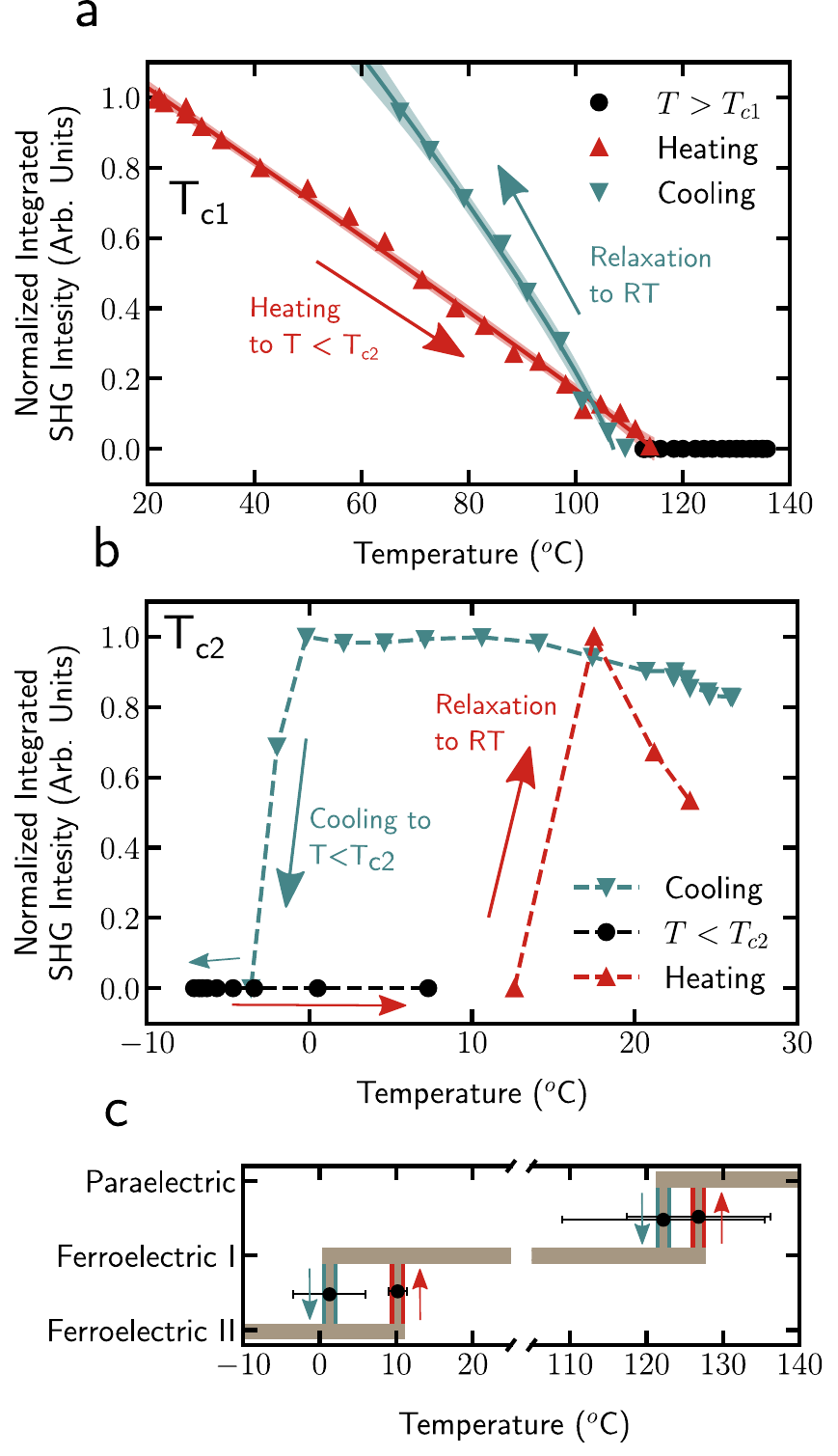}
\caption{\label{fig:transition}SHG intensity as a function of temperature for each phase transition and its subsequent relaxation. \textbf{a} The tetragonal-cubic phase transition can be fitted with a Curie-Weiss-law as shown in Equation~(2) and yields critical exponents $\beta$ close to the ideal value for a second-order transition. \textbf{b} In contrast, the orthorhombic-tetragonal phase transition shows a sudden jump in intensity while cooling through the transition temperature, as well as during relaxing indicating a first-order phase transition. In both graphs, a pronounced hysteresis behavior is visible. Subfigure~\textbf{c} shows the phase diagram of the investigated sample. The data points in \textbf{c} are averaged over multiple experiments.}
\end{figure}

According to Ginzburg-Landau-Devonshire theory, the evolution of the order parameter $P_S$  as a function of temperature $T$ of a ferroelectric is described by a Curie-Weiss law of the form

\begin{equation}
P_S(T) \propto  \left( 1 - \frac{T}{T_C} \right)^{\beta}
\end{equation}

where $\beta$ is the critical exponent. For an archetypal second-order-transition it holds $\beta = 1/2$. Any deviations from this value can indicate, for example, disorder or mixed type transitions. For example, for strontium barium niobate (SBN), which is a relaxor and not a classical ferroelectric, values of $\beta_{SBN} \approx 0.13$ have been observed \cite{Volker2006,Ayoub17}.

The second-harmonic generation intensity $I_{SHG}$ of any material is proportional to the square of the nonlinear susceptibility of the material, which in turn is directly proportional to the spontaneous polarization. Therefore, it holds

\begin{equation}
I_{SHG} \propto \left( P_S(T) \right)^2 \propto  \left( 1 - \frac{T}{T_C} \right)^{2 \beta},
\end{equation}

which can directly be used to extract $\beta$ from the integrated SHG intensity as a function of temperature \cite{Volker2006,Ayoub11,Ayoub17}.

Therefore, to extract the critical exponent from our datasets, the SHG intensity as a function of temperature needs to be evaluated. For this, we simply calculate for each image in a cooling or heating cycle the total SHG intensity by integrating the intensities from all pixels, which results a single total intensity value for each image.  This value is then evaluated as a function of temperature. It should be noted, in our case the SHG intensity is mostly emitted from the domain walls by the \v{C}erenkov-type phase matching and from the bulk. As we see below, the intensity emitted by the DWs indeed scales in accordance with Equation (2), which was also observed in other materials \cite{Ayoub11,Ayoub17}.

Figure~\ref{fig:transition} shows an exemplary data set during \textbf{a} the cubic-tetragonal and \textbf{b} the tetragonal-orthorhombic transitions, respectively. Here, each dot datapoint represents the intensity of one SHG image. Because we are only interested in the evolution of intensity as a function of temperature, but not the absolute intensity, the intensity was normalized with respect to \textbf{a} the intensity at room temperature before the heating cycle and \textbf{b} the maximum intensity observed just before the phase transition. To enhance clarity in each data set, the data-points are further sorted for the heating, cooling, and relaxation phases, as well as the data points below or above the respective phase transition temperature. For these experiments, the focus was placed at a depth of 50~\textmu m instead of 5~\textmu as shown above. The raw-data sets, i.e., intensity and temperature as a function of time, and some domain structure images from this depth are shown in the supplementary information (Figs.~S1 and S2). The frame rate was larger than 1~Hz, while the cooling/heating rate was less than 1~K/s.

Fitting the data set of the tetragonal-cubic phase transition for heating above $T_{c1}$ and relaxation back to room temperature (Fig.~\ref{fig:transition}\textbf{a}) using Eq.~(2), respectively, yields a critical exponent $\beta = 0.48 \pm 0.02$ during heating and $\beta = 0.43 \pm 0.02$ during relaxation to room temperature, which is very close to the expected value (0.5) for a pure second-order transition. Please note on relaxation the total intensity is larger compared to before the phase transition. This is indicative of the increased number of DWs. However, for evaluating the critical exponent via Equation~(2) only the shape and slope of the SHG-intensity-vs.-temperature characteristics, but not the total intensity is relevant. 

In this experimental run, a pronounced hysteresis is observed, with $T_{c1} = 116$~$^\circ$C during heating, but at relaxation we observed a transition temperature as low as $T_{c1} = 107$~$^\circ$C. This is lower compared to the qualitative observations above. A possible explanation is that the data set displayed in Fig.~\ref{fig:transition} is from the very first experimental run on this sample, while in later experimental runs similar critical exponents, but a higher transition temperature is observed (see Fig.~\ref{fig:domains} and Table~S1 in the Supplementary Material). Here, the very first heating cycle represents a quasi-annealing experiment, after which the sample is in a more stable configuration leading to more reproducible temperatures. 

Furthermore, it is observed that during relaxation the critical exponent is slightly smaller, while the overall intensity recovers to a higher value. This may be indicative of the larger density and different patterning of domains we observed during such cycles, i.e., the annealing above $T_{c1}$. Our observations of a gradual change of SHG intensity point towards a second-order phase transition and are similar to previous macroscopic SHG studies of this phase transition \cite{Wang2018,Pugachev2012}. In our methodology, however, we can observe this behavior while also imaging the domain structure.

Figure~\ref{fig:transition}\textbf{b} shows a data set during cooling through the orthorhombic-tetragonal phase transition and subsequent relaxation. Here, in contrast to Fig.~3\textbf{a} the total SHG intensity emitted from the investigated area stays approximately constant, until a sudden drop in intensity is observed coinciding with the apparent disappearance of the (tetragonal) domain structure. Within the limits of the accuracy of the experiment this behavior can not be fitted with  Equation~(2) and appears to be drastically different from the tetragonal-cubic phase transition, which is in agreement with previous observations with PFM revealing this phase transition to be of first-order nature \cite{Limboeck2014,doe18}. The plot also highlights the strong hysteresis when relaxing back to room temperature, which happens at an almost 15~K higher temperature as compared to the phase transition upon cooling.

Some PFM studies\cite{doe18,Limboeck2014} reported a "mixed phase", i.e., a continuous reappearing of the tetragonal domain structure, between +8$^{\circ}$C and +16$^{\circ}$C  during heating from the orthorhombic to the tetragonal phase. In contrast, in the present study, the transition back into the tetragonal phase was not found to be continuous, but rather sudden. One reason for this might be that the temperature was adjusted by gradients of about 1~K/s rather than e.g.~1~K/min, as was reported in other works. In future work, lower temperature gradients are thus desired to investigate the possible existence of such a mixed phase. An alternative possibility is an effect of the stoichiometry and crystal quality, i.e. defects, which may be different in our study compared to previous work \cite{doe18,Limboeck2014}.

Apart from determining the critical exponent $\beta$, the fitting procedure also yields the Curie-temperature. Here, in general, the fit provides a slight overestimate compared to the temperatures obtained by determining the first SHGM image in the series, where no SHG intensity is observed. As seen above, the fit procedure provided no useful results for the orthorhombic-tetragonal phase transition. Therefore, to obtain overall comparable values  for $T_{c1}$ and $T_{c2}$, temperatures have been extracted by directly determining the temperature, where the SHG intensity is first lost or observed after relaxation, respectively. 

Figure~\ref{fig:transition}\textbf{c} shows the determined phase diagram by averaging over multiple experiments. As expected, we observe for both transitions a pronounced hysteresis, as well as a broad range of transition temperature values. During multiple experimental runs, we observed values of $T_{c1} \approx 120-130$~$^\circ$C with the lowest value being $T_{c1} = 116$~$^\circ$C and the largest being $T_{c1} = 133$~$^\circ$C. During relaxation we observe values as low as $107$~$^\circ$C, but on average a range of $T_{c1} \approx 110-125$~$^\circ$C. For the orthorhombic-tetragonal phase transition we observe initial transition temperatures of $T_{c2} \approx 0-5 $~$^\circ$C, which are comparable ranges commonly reported in literature\cite{Limboeck2014,doe18,Villafuerte2016,MAZUR2023}. The relaxation temperature of $\approx 14 $~$^\circ$C is very well in agreement with previous reports.  In particular, the lowest temperatures are observed during the first experimental runs, which indicate signs of annealing by the experiments themselves. 

\section{Conclusion}

In this work, we have investigated the paraelectric-ferroelectric I (cubic-tetragonal) and ferroelectric I-ferroelectric II (tetragonal-orthorhombic) phase transitions in the archetypal ferroelectric crystal BaTiO$_3$ with SHGM, which made it possible (i) to directly extract the transition temperatures of both transitions, (ii) to visualize the impact of temperature sweeps on the domain structure in an arbitrary depth of the crystal, as well as (iii) to extract the critical exponent (thermodynamic order) of the phase transition from the behavior of the SHG intensity emitted from the domain walls. Here, we found -- in agreement with previous research by PFM or PLM -- the paraelectric-ferroelectric I (cubic-tetragonal) at a transition temperature of $T_{c1} \approx 120-130$~$^\circ$C and the ferroelectric I-ferroelectric II (tetragonal-orthorhombic) phase transition at a range of $T_{c1} \approx 0^\circ - 5$~$^\circ$C. We observed that the overall shape of the domain structure usually recovers after relaxation through the tetragonal-orthorhombic phase transition indicating the important role of defect pinning centers for the natural, local domain structure. In contrast, after heating and relaxation through the cubic-tetragonal transition, the newly achieved domain structure appears to be more homogeneous. Due to the 3D resolution of SHGM this can be confirmed also within the bulk of the crystal. Furthermore, the paraelectric-ferroelectric I (cubic-tetragonal) transition was found to be second-order with a critical exponent $\beta = 0.48 \pm 0.2$ within errors coinciding with the ideal value of 1/2. In contrast, the drastic and sudden change of intensity in the ferroelectric I-ferroelectric II (tetragonal-orthorhombic) strongly indicated a first-order behavior. In our experiment, this was determined by evaluating the \v{C}erenkov-type SHG intensity emitted from the domain walls, which demonstrated that also the \v{C}erenkov-type SHG signal from the domain walls behaves similar to bulk SHG during the phase transition \cite{Wang2018,Pugachev2012}. Due to its ease of application requiring no special sample preparation, its non-destructive nature, and direct sensitivity to domain structures, SHGM can be readily extended to other ferroelectric materials or other ferroic materials with associated nonlinear properties. 



\section*{Acknowledgements}

We acknowledge financial support by the Deutsche Forschungsgemeinschaft (DFG) through joint DFG--ANR project TOPELEC (EN~434/41-1 and ANR-18-CE92-0052-1), the CRC~1415 (ID: 417590517), the FOR~5044 (ID: 426703838; \url{https://www.for5044.de}), as well as through the W\"urzburg-Dresden Cluster of Excellence on "Complexity and Topology in Quantum Matter" - ct.qmat (EXC 2147, ID: 39085490). This work was supported by the Light Microscopy Facility, a Core Facility of  the CMCB Technology Platform at TU Dresden. I.K.'s contribution to this project is also co-funded by the European Union and co-financed from tax revenues on the basis of the budget adopted by the Saxon State Parliament.






\section*{References}

\bibliography{literature}

\beginsupplement

\onecolumngrid
\newpage
\section*{Supplementary Information}

\begin{figure*}[ht]
\includegraphics[width=0.8\textwidth]{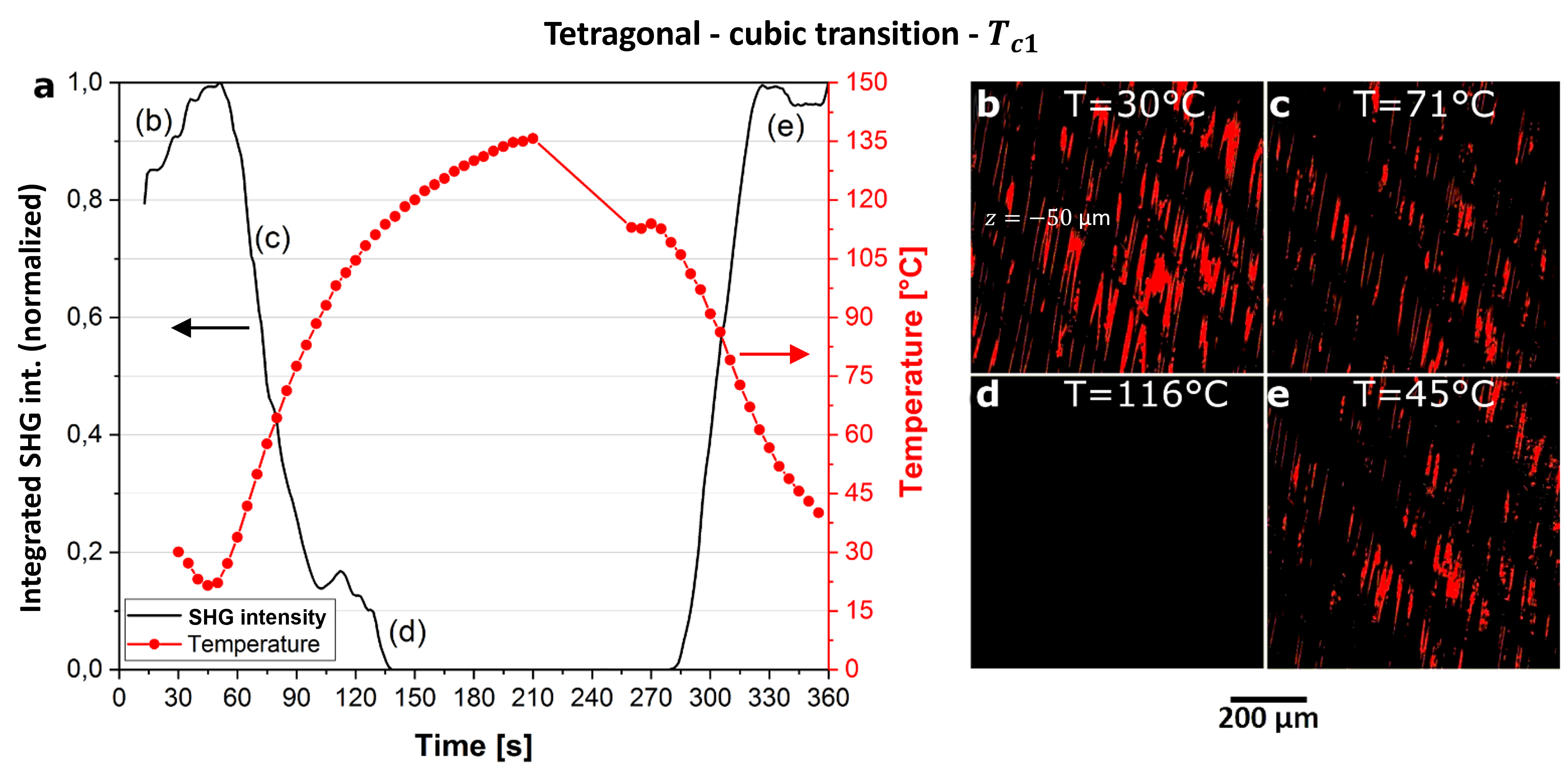}
\caption{\label{fig:depth_data_tc1} Heating cycle above the tetragonal-cubic phase transition of the investigated sample in a depth of 50~\textmu m. Panel \textbf{a} shows the integrated SHG intensity (left axis) and the temperature evolution (right axis) as a function of time. The same data is evaluated for extracting the critical exponent (Fig.~3\textbf{a} in the main text). Panels \textbf{b} to \textbf{e} shows SHGM pictures in a depth of 50~\textmu m at selected temperatures. The pattern of domains and behavior are similar to the previous experiments at a depth of 5~\textmu m. In this depth a pattern of lamella-like 180$^\circ$C. The oscillation pattern visible in the data likely comes from some form of thickness-related interference and was observed in other materials before \cite{Amber2022,amber2021quantifying,Spychala2023} and is not related to the domain structure.}
\end{figure*}

\begin{figure*}[ht]
\includegraphics[width=0.8\textwidth]{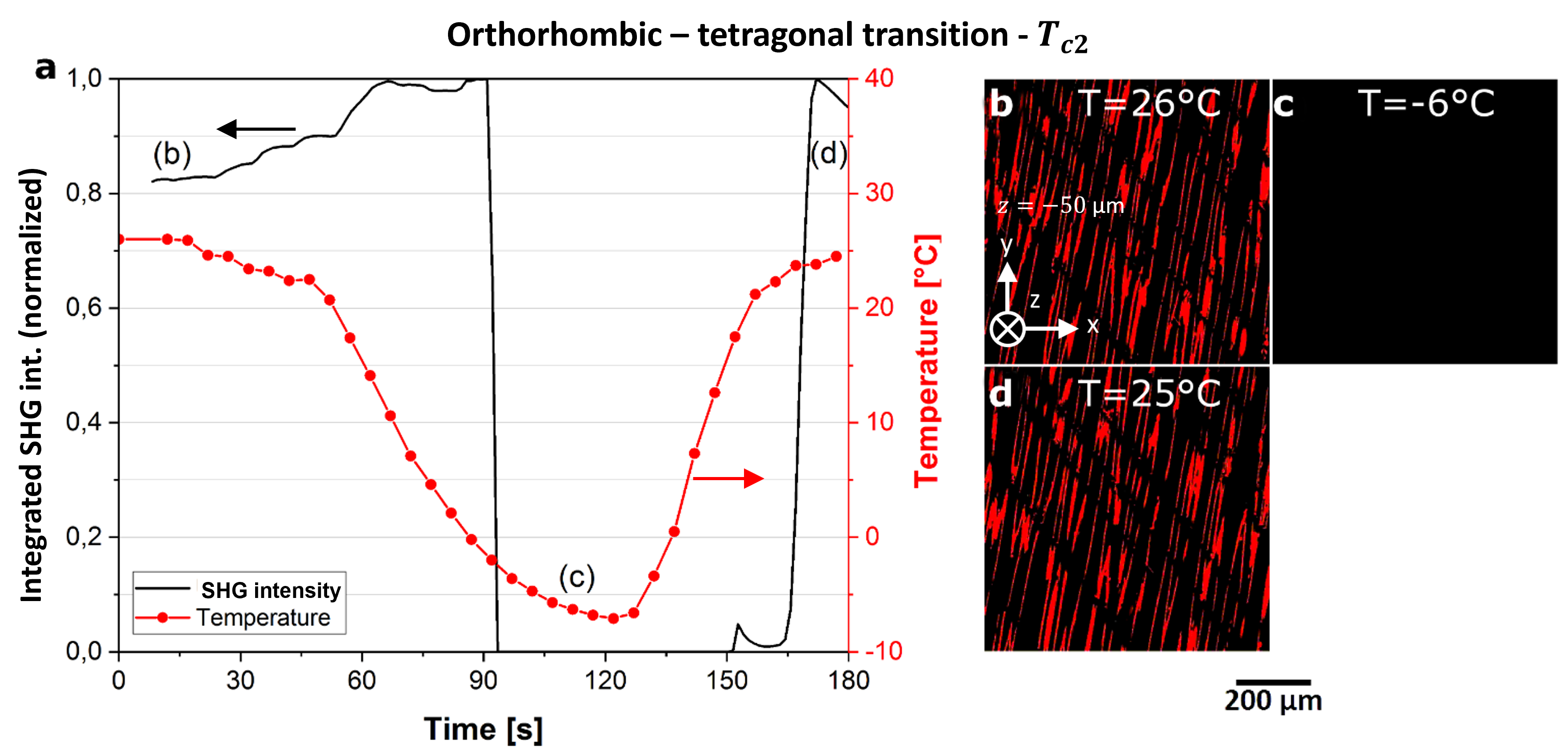}
\caption{\label{fig:depth_data_tc2} Cooling cycle below the tetragonal-orthorhombic phase transition of the investigated sample in a depth of 50~\textmu m. Panel \textbf{a} shows the integrated SHG intensity (left axis) and the temperature evolution (right axis) as a function of time. The same data is evaluated for extracting the critical exponent (Fig.~3\textbf{b} in the main text). Panels \textbf{b} to \textbf{e} shows SHGM pictures in a depth of 50~\textmu m at selected temperatures. The pattern of domains and behavior are similar to the previous experiments at a depth of 5~\textmu m. In this depth a pattern of lamella-like 180$^\circ$C. The oscillation pattern visible in the data likely comes from some form of thickness-related interference and was observed in other materials before \cite{Amber2022,amber2021quantifying,Spychala2023} and is not related to the domain structure.}
\end{figure*}

\begin{table*}
\centering
\caption{Measured transition temperatures $T_{c1}$ and $T_{c2}$ during cooling, heating and relaxation from/to room temperature. For each of the transitions two values are given, one corresponding to the transition temperature observed during heating or cooling, while the second value corresponds to the transition temperature during relaxation.}
\vspace{10mm} 
\begin{tabular}{cccc}

\hline
\multicolumn{2}{c}{$T_{c2} [\symbol{23} C]$} & \multicolumn{2}{c}{$T_{c1} [\symbol{23} C]$} \\
cooling & relaxation & heating & relaxation  \\
\hline
 -2.0 & 13.0 & 107.0 & 116.0  \\
4.6  & 11.0 & 129.6 & 133.0 \\
      &      & 130.1 & 131.3 \\
\hline
\end{tabular}

\label{Tab_BTO_Curie}
\end{table*}

%

\end{document}